\documentclass{cargese}
\usepackage{graphicx}
\usepackage{rotating}
\usepackage{amssymb}
\usepackage{pstricks}

\let\footnote\savefootnote
\let\footnotetext\savefootnotetext
\let\footnoterule\savefootnoterule
\normallatexbib

\def\dalemb#1#2{{\vbox{\hrule height .#2pt
        \hbox{\vrule width.#2pt height#1pt \kern#1pt
                \vrule width.#2pt}
        \hrule height.#2pt}}}

\def\tA{\widetilde A}

\def\0{{\sst{(0)}}}
\def\1{{\sst{(1)}}}
\def\2{{\sst{(2)}}}
\def\3{{\sst{(3)}}}
\def\4{{\sst{(4)}}}
\def\5{{\sst{(5)}}}
\def\6{{\sst{(6)}}}
\def\7{{\sst{(7)}}}
\def\8{{\sst{(8)}}}

\def\tF{\widetilde F}
\def\tA{\widetilde A}

\def\G{{\cal G}}

\def\S{{\cal S}}

\def\nn{\nonumber} \def\bd{\begin{document}} \def\ed{\end{document}}
\def\ds{\documentstyle} \let\fr=\frac \let\bl=\bigl \let\br=\bigr
\let\Br=\Bigr \let\Bl=\Bigl
\let\bm=\bibitem
\let\na=\nabla
\let\pa=\partial \let\ov=\overline
\newcommand{\be}{\begin{equation}}
\newcommand{\ee}{\end{equation}}
\def\ba{\begin{array}}
\def\ea{\end{array}}
\def\ft#1#2{{\textstyle{{\scriptstyle #1}\over {\scriptstyle #2}}}}
\def\fft#1#2{{#1 \over #2}}
\def\del{\partial}
\def\sst#1{{\scriptscriptstyle #1}}
\def\oneone{\rlap 1\mkern4mu{\rm l}}
\def\ie{{\it i.e.\ }}
\def\via{{\it via}}
\def\semi{{\ltimes}}
\def\v{{\cal V}}
\def\str{{\rm str}}
\def\Dm{{{D_{\sst{max}}}}}

\newcommand{\ho}[1]{$\, ^{#1}$}
\newcommand{\hoch}[1]{$\, ^{#1}$}
\newcommand{\bea}{\begin{eqnarray}}
\newcommand{\eea}{\end{eqnarray}}
\newcommand{\ra}{\rightarrow}
\newcommand{\lra}{\longrightarrow}
\newcommand{\Lra}{\Leftrightarrow}
\newcommand{\bp}{\tilde \beta^\prime}
\newcommand{\tr}{{\rm tr} }
\newcommand{\Tr}{{\rm Tr} }

\newcommand{\CP}{{\mathbb P}}

\begin{document}

\articletitle[Symmetries in M-theory: \\ MONSTERS, INC.]
{Symmetries in M theory: \\  MONSTERS, INC. }

\chaptitlerunninghead{Symmetries in M-theory: MONSTERS, INC.}
\author{P.
Henry-Labord\`ere$^1$\footnote{Talk given by PHL at Cargese 2002.}, B. Julia$^2
$, L. Paulot$^2$ }



\affil{
1: Queen Mary and Westfield College, University of London \\
Mile End Road, London E1 4NS
\\
2: Laboratoire de Physique th\'eorique de l'Ecole Normale Sup\'erieure,
24 rue Lhomond, 75231 Paris Cedex 05, France.}

\email{p.henry-labordere@qmul.ac.uk}
\email{bernard.julia@lpt.ens.fr}
\email{paulot@lpt.ens.fr}

\begin{abstract}
We will review the algebras which have been conjectured as symmetries in M-
theory. The Borcherds algebras, which are the most general Lie algebras under
control, seem  natural candidates.
\end{abstract}

Little is known about  M-theory except that its low effective
action is described by eleven-dimensional supergravity
\cite{jul1}. The dimensional reduction of this supergravity theory
on a $n$-dimensional torus $T^n$ was shown to be invariant under
the split form $E_{(n|n)}$ of the complex exceptional Lie algebra
$E_n$ \cite{jul2}. The split form of a complex Lie algebra is
defined by the restriction of the field of coefficients from
complex to real numbers and in M-theory, it has been
conjectured that we must replace the real field by the ring of integers
in the Cartan Weyl basis.
\cite{hul}. These arithmetic groups, called U-dualities, have been
extensively used to compute some non-perturbative contributions in
string theory. The well-known example is the $t_8t_8R^4$
term in
$IIB$ superstring which can be computed exactly using
supersymmetry
constraints and the modular group $SL(2,{\mathbb
Z})$. The solution involves  non-holomorphic Eisenstein
forms.

Using a rich connection between particular complex algebraic surfaces, called
del Pezzo surfaces, and field theories, extending the work \cite{vaf} presented by A.
Neitzke in this volume, we have shown that the U-duality algebras can be
enlarged into some {\it Borcherds (super)algebras} \cite{pbl1}.  This class of
algebras, which contains the Kac-Moody algebras, was introduced by Borcherds
\cite{bor} in order to prove the Moonshine conjecture which states
that the characters of  representations of a sporadic group, called the
Monster, are modular forms under  Hecke subgroups of $SL(2,{\mathbb Z})$.

In our correspondence, the  simple roots of the Borcherds algebra are
associated to a del Pezzo surface $X^{{\mathbb C}}$, they span the full Picard group
$Pic(X^{{\mathbb C}})$ and the cone containing the  positive roots is defined
as the convex hull of the rational
(i.e. of vanishing virtual genus) divisor classes of non negative degree
(one must also exclude one half of the degree zero hyperplane).
We can then read off the intersection matrix
$A_{ij}=\alpha_i.\alpha_j$ and a ${\mathbb Z}_2$-graduation by $grad(\alpha_i)
=-K.\alpha_i \, \, mod \, 2$. However, it turns out that whenever a fermionic
root of square $-1$ appears it should be viewed as an $sl(1|1)$ superroot,
i.e. it should have zero (intersection)-Cartan-Killing norm.
The corresponding modified intersection matrix will be minus our
Cartan matrix $a_{ij}$.
This symmetric matrix $a_{ij}$ satisfies the defining properties
of a  Borcherds superalgebra:
\begin{eqnarray}
(i) & a_{ij} \leq 0 &
\mathrm{if} \ i \neq j \\
(ii) & \frac{2a_{ij}}{a_{ii}} \in  {\mathbb Z} &
\mathrm{if} \
a_{ii} > 0
\ ,\
\mathrm{for} \ grad(\alpha_i)=0
\\
(iii) & \frac{a_{ij}}{a_{ii}} \in {\mathbb Z}
&
\mathrm{if} \ a_{ii}
> 0 \
, \ \mathrm{for} \ grad(\alpha_i)=1
\end{eqnarray}

\begin{figure}
\begin{center}
\label{dyn}

\begin{picture}(100,80)
\multiput(0,40)(20,0){3}{\circle{10}}
\put(0,25){\makebox(0.4,0.6){ $\alpha_1$}}
\put(20,25){\makebox(0.4,0.6){$\alpha_2$}}
\put(40,25){\makebox(0.4,0.6){ $\alpha_3$}}
\put(100,25){\makebox(0.4,0.6){$\beta$}}
\put(40,60){\circle{10}}
\put(60,40){\dots}
\put(100,40){\circle{10}}
\put(40,70){\makebox(0.4,0.6){ $\alpha_0$}}
\put(40,45){\line(0,1){10}}
\put(95,40){\line(-1,0){10}}
\put(5,40){\line(1,0){10}}
\put(25,40){\line(1,0){10}}
\put(45,40){\line(1,0){10}}
\end{picture}

\caption{Dynkin diagram.}
\end{center}
\end{figure}
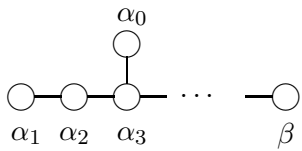

The Borcherds superalgebra associated to the matrix $a_{ij}$ is generated by
its Cartan subalgebra, the positive and negative generators satisfying the
Chevalley-Serre (and superJacobi) relations. The Dynkin diagram of the real
split
Borcherds superalgebra corresponding to eleven-dimensional supergravity
compactified on a $n$-torus ($n>3$) is given in the figure \ref{dyn}. $\beta$
is a fermionic simple root of vanishing norm and $\alpha_i$ are  simple
bosonic
roots of norm 2. The roots $(\alpha_i,\beta)$ for $i\geq 1$
define a sl($n \arrowvert 1$)
superalgebra and the roots $(\alpha_i)$ define the Dynkin diagram of the U-
duality group $E_{(n|n)}$.

In \cite{pbl2}, we have shown that the non-split form of
these Borcherds superalgebras, which appear in supergravity theories with
less than $32$ supersymmetries, correspond also to possibly singular
real del Pezzo surfaces.
The real structure of the surface is a conjugation which preserves the intersection product and the canonical
class $-K$, it is identified with the conjugation $\sigma$ defining  the real
form of the Borcherds algebra and
defines a {\it Satake superdiagram}, which is a bicoloured Dynkin diagram with
black vertices for the anti-invariant roots under $\sigma$.

Note that the Borcherds (super)algebras have already appeared in
superstring theory \cite{hm1,dij,kaw,bor1}. Indeed, the physical states of a
(super)string fully compactified on a Lorentzian lattice define a Borcherds
(super)algebra. For example, if we compactify the superstring on the torus
$T^{9,1}$ we obtain the {\it Fake Monster superalgebra} which contains the
root lattice of $E_{10}$  and a ${\mathbb Z}_2$-orbifolding gives the {\it
Monster superalgebra} \cite{nic,sch}. It is important to observe that if we
compactify the heterotic string on a torus the Narain lattice is not
generally Lorentzian; the spectrum of perturbative $1/2$-BPS states defines
a vertex algebra and forms a {\it Generalized Borcherds algebra} which has not
been studied in Mathematics \cite{hm1}. The S-dual Lie algebra of the
perturbative $1/2$-BPS states on $T^4$, corresponding to
$1/2$-BPS D-branes of $IIA$ on $K3$, can be obtained as the vertex algebra
associated to the Picard group of $K3$. This is called the Nakajima
construction \cite{nak}.

Recently, $E_{10}$ (and $E_{11}$), introduced by one of us as candidate
U-duality groups of
eleven-dimensional supergravity compactified to one (and zero dimension), have
been conjectured to be symmetries of M-theory in eleven dimension \cite
{dam1,wes}. Actually the Weyl group of $E_{10}$ appears when one studies the
chaotic behaviour of M-theory near a cosmological singularity. The oscillatory evolution of generic
cosmological solutions of Einstein's D-dimensional gravity with several other
fields coming from string theory or
M-theory can be approximated by a flow in a relativistic billard defined by
the fundamental
Weyl chamber of some Kac-Moody algebra. The solutions are chaotic if the
corresponding Kac-Moody algebra is hyperbolic \cite{dam2}.  All these algebras
suggest a new formulation of M-theory which should incorporate the fields and
their duals in a symmetric way \cite{cjlp}.

In the next part of this review, we will show that the classical equation of
motion for the 3-form $A_3$ in M-theory \cite{cjlp} can be
obtained using the superalgebra $osp(1|2)$ \cite{pbl1}, which is generated by
the following relations for the positive generators:
\begin{equation}
\{e_{\alpha_0},e_{\alpha_0}\}= -e_{\alpha_1}\ ,\qquad
{[}e_{\alpha_0},e_{\alpha_1} {]}=0\ ,\qquad {[} e_{\alpha_1}, e_{\alpha_1}{]}
=0\ .
\label{d11com}
\end{equation}
Let us introduce the following nonlinear ``potential'' differential form:
\be
\v = e^{A_\3\, e_{\alpha_0}}\, e^{\tA_\6\,e_{\alpha_1} }\ .\label{d11coset}
\ee
The Grassmann {\it angle} $A_\3$ (resp. $\tA_\6$) is a 3-form (resp. 6-form)
coupled to the M2- (resp. M5-) brane  and
defined on an eleven dimensional manifold $X$.

By an elementary calculation one checks that the field strength $\G=d\v\,
\v^{-1}$, satisfying the Maurer-Cartan equation $d\G=\G \wedge \G$, following
from (\ref{d11coset}) is given by
\bea
\G &=& (dA_\3) \,e_{\alpha_0}  + (d\tA_\6 - \ft12 A_\3 \wedge dA_\3)
\,e_{\alpha_1} \ ,\nn\\
  &=& F_\4  \, e_{\alpha_0} + \tF_\7\, e_{\alpha_1}\ .\label{gcalc}
\eea
Now, we introduce the self-duality equation: $ \S \G = * \G$ where $\S$ is an
operator which exchanges $e_{\alpha_0}$ with $e_{\alpha_1}$. Using (\ref
{gcalc}), we obtain  \be *F_\4=F_\7 \label{eom} \ee
By taking the exterior derivative of this equation, we obtain the equation of
motion of the 3-form in eleven-dimensional supergravity. This equation can be
shown to be invariant under the Borel
subgroup of a supergroup $OSP(1|2)$. Indeed, an element $\Lambda$ of this
Borel subgroup satisfying $d\Lambda=0$ acts on $\v$ on the right:
$\v'=\v.\Lambda$. The field strength is unchanged and the self-duality
equation is preserved.

This construction can be generalized for the other toroidal compactifications
of that theory and finite subsuperalgebras of the above Borcherds
superalgebras preserve
the equations of motion for the various $p$-forms in supergravity theories.

In \cite{llps} and recently in \cite{ste}, it has been shown that the
superalgebra
(\ref{d11com}) implies nonlinear relations of the type
$\, {t_{\alpha_0} \over 2 \pi}.{t_{\alpha_0} \over 2\pi}={t_{\alpha_1}
\over 2\pi} \,$ where $t_{\alpha_0}
$ (resp. $t_{\alpha_1} $) is
the tension of the M2 brane (resp. M5 brane). Then using Dirac's
quantization condition the tensions can be computed.
These relations can be generalized for our Borcherds superalgebras. This
analysis suggest also that our Borcherds superalgebras must be broken in M-
theory into an arithmetic supergroup with angles given by forms in $H^*(X,
{\mathbb Z})$.



\begin{acknowledgments}
P. Henry-Labord\`ere would like to thank the organizers of the Carg\`ese 2002
ASI for giving him the opportunity to present
this work \cite{pbl1}.

\end{acknowledgments}


\end{document}